\documentclass[10pt]{article}
\usepackage{graphicx}
\usepackage{txfonts}

\begin{document}
\title{Non-Commutative Geometry and Measurements of
Polarized Two Photon Coincidence Counts}
\author{S. Sivasubramanian, G. Castellani, N. Fabiano$^\dagger $, \\
A. Widom, J. Swain, Y.N. Srivastava$^\dagger $,
G. Vitiello$^{\dagger \dagger}$ \\
\\
Physics Department
Northeastern University, Boston MA USA \\
$^\dagger $Physics Department \& INFN,
University of Perugia, Perugia Italy \\
$^{\dagger \dagger}$ Physics Department \& INFN,
University of Salerno, Salerno Italy}
\date{}
\maketitle
\begin{abstract}
Employing Maxwell's equations as the field theory of the photon,
quantum mechanical operators for spin, chirality, helicity, velocity,
momentum, energy and position are derived. The photon ``Zitterbewegung''
along helical paths is explored. The resulting non-commutative geometry
of photon position and the quantum version of the Pythagorean theorem
is discussed. The distance between two photons in a polarized beam
of given helicity is shown to have a discrete spectrum. Such a
spectrum should become manifest in measurements of two photon
coincidence counts. The proposed experiment is briefly described.
\end{abstract}

\section{Introduction}

In recent years there has been considerable mathematical
and physical interest in non-commutative geometry\cite{Nair:2001}
and its
physical consequences. Perhaps, the most simple example of a non-commutative
geometry consists of the geometric plane described by Hermitian operator 
coordinates
\begin{math} (\hat{X},\hat{Y})  \end{math}
that do not commute; e.g. there exists an area
\begin{math} {\cal A} \end{math} such that
\begin{equation}
i\left[\hat{X},\hat{Y}\right]={\cal A}.
\label{intro1}
\end{equation}
As a consequence of Eq.(\ref{intro1}), there is a ``quantization'' of
the Pythagorean theorem in that the distance
\begin{math} \hat{D}=\sqrt{\hat{X}^2+\hat{Y}^2} \end{math}
has a quantized spectrum\cite{Sivasubramanian:2003xy}. For
\begin{math} n=0,1,2,3,\ldots
\end{math}\ , we have
\begin{equation}
(\hat{X}^2+\hat{Y}^2)\left|n\right> = D_n^2\left|n\right>
\ \ {\rm where}\ \ D_n=\sqrt{(2n+1){\cal A}}\ .
\label{intro2}
\end{equation}

Experimental examples of physical systems which can be usefully
described by coordinates in the non-commutative plane include
the following: (i) vortices in superfluid
\begin{math} ^4He  \end{math} films\cite{Yourgrau:1968},
(ii) quantum Hall effect charged magnetic vortices in
two-dimensional electron liquids\cite{Belissard:1988}, (iii)
quantum interference phase between two alternative paths in the
plane (as in the Aharonov--Bohm
effect)\cite{Sivasubramanian:2003xy} and (iv)
 high energy charged
lepton beams stored in
cyclotrons\cite{Widom:2001,Srivastava:2002,Y1}. In all of these
experimental examples, there exists the ordinary Euclidean
geometry of position vectors
\begin{math} ({\bf r}_1, {\bf r}_2,\ldots )\end{math} of the particles
which make up the system. The Euclidean coordinates {\em may} appear
in the quantum mechanical wave functions of the system. Euclidean
coordinates may also appear in the classical rulers employed to construct
the lengths scales of the apparatus used in measuring the quantum mechanical
system properties. However, the quantum coordinates
(such as vortex positions or the positions of cyclotron
orbit centers) are described by strictly non-commuting operators. Thus,
both classical Euclidean geometry and quantum non-commutative geometry
live tranquilly together in the theoretical description of the
above experimental quantum systems.

Such tranquility also exists in the quantum mechanical behavior of zero
mass non-zero spin quantum mechanical particle beams. These include
neutrinos, photons\cite{Wigner:1939} and gravitons. Our purpose is to
discuss the case of zero mass {\em photons}
\cite{Pryce:1948,Wigner:1949,Wightman:1962,Juach:1967,Hawton:1998,Schwinger:1998},
where the experimental technology\cite{Mandel:1995} (of quantum optics)
is most fully developed. Our central theoretical results concern two photon
coincidence counts measured from polarized beams. The statistics of the counts
should reflect the non-commutative geometry of the photon positions.

In Sec.\ref{me} we write Maxwell's equations in the vacuum in a form 
conducive to describing the photon as a spin one object. The quantum 
mechanical spin, chirality, helicity, velocity, momentum and energy 
operators for a single photon will be derived. In Sec.\ref{pv} 
the ``velocity'' operator of the
photon will be explored. Writing the photon velocity operator as
\begin{math} \dot{\bf r}={\bf v}=(v_x,v_y,v_z)  \end{math},
one finds that each component {\em formally} has the possible eigenvalues
\begin{math} -c,\ 0\ {\rm or}\ +c \end{math}. However,
each component also has non-trivial commutation relations with the other
components finally yielding
\begin{equation}
{\bf v\cdot v} = v_x^2+v_y^2+v_z^2=2c^2>c^2\ \ ({\rm photons}).
\label{intro3}
\end{equation}
Eq.(\ref{intro3}) indicates a considerable amount
of ``Jitterbugging'' in the photon motion. For a Dirac particle, wherein
\begin{math} {\bf v}_{Dirac}\cdot {\bf v}_{Dirac}=3c^2 \end{math},
Schr\"odinger called the resulting (superluminal) velocity fluctuations
{\em Zitterbewegung} (German Jitterbugging). In more detail, it will be shown
that the component of photon velocity perpendicular to the photon momentum
\begin{math} {\bf p} \end{math} undergoes a rotation at angular velocity
\begin{math} \Omega =c|{\bf p}|/\hbar  \end{math}. The resulting photon 
``path'' may then be described as {\em helical}. Along the axis of the 
helix, the photon moves with velocity \begin{math} c \end{math}. 
But the photon also moves around the helix in such a way that the total 
speed is greater than \begin{math} c \end{math}.
In Sec.\ref{po}, the Wigner photon position operator will be discussed.
The photon position will be shown to obey the rules of a non-commutative
geometry with an associated quantum Pythagorean theorem. The motivation for
introducing such a position operator can be understood in terms of the photon
angular momentum as is discussed in Sec.\ref{am}.
In Sec.\ref{pa}, the non-commutative geometry of a two photon system will be
discussed. The distance between two photons in a polarized beam is shown 
to have a discrete spectrum. Such a spectrum should become manifest in 
measurements of two photon coincidence counts. In the concluding 
Sec.\ref{co}, the technology
of these proposed experiments will be briefly described.

\section{Maxwell's Equations \label{me}}
The Maxwell equations describing a photon in the vacuum are well known. The
relativistic covariance is well established so that we pick a particular
inertial frame (say the ``laboratory frame'') in which they read
\begin{eqnarray}
div {\bf E}&=&0,
\nonumber \\
div {\bf B}&=&0,
\nonumber \\
curl{\bf E}&=& -\frac{1}{c}\left(\frac{\partial {\bf B}}{\partial t}\right),
\nonumber \\
curl{\bf B}&=& \frac{1}{c}\left(\frac{\partial {\bf E}}{\partial t}\right).
\label{me1}
\end{eqnarray}
Introducing the complex electromagnetic fields 
\begin{math} {\bf F}_+ \end{math}
and \begin{math} {\bf F}_- \end{math} via
\begin{equation}
{\bf F}_\pm = {\bf E} \pm i{\bf B},
\label{me2}
\end{equation}
allows us to write Maxwell's equations in the more compact form
\begin{equation}
div{\bf F}_\pm =0\ \ \ {\rm and}
\ \ \ i\left(\frac{\partial {\bf F}_\pm }{\partial t}\right)
=\pm c\ curl{\bf F}_\pm .
\label{me3}
\end{equation}
The operator {\it curl\ } defines the spin one matrices
\begin{math} {\bf S}=(S_x,S_y,S_z) \end{math} according to the
equivalent definitions
\begin{equation}
{\bf U} = curl{\bf W}
\label{me4}
\end{equation}
and
\begin{equation}
\pmatrix{U_x \cr U_y \cr U_z} =
\pmatrix{0 & -\partial_z & \partial_y\cr
\partial_z & 0 & -\partial_x\cr
-\partial_y & \partial_x & 0}
\pmatrix{W_x \cr W_y \cr W_z}=-i({\bf S\cdot grad })
\pmatrix{W_x \cr W_y \cr W_z}.
\label{me5}
\end{equation}
Eq.(\ref{me5}) implies the explicit representation
\begin{equation}
S_x=\pmatrix{0 & 0 & 0 \cr 0 & 0 & -i\cr 0 & i & 0}, \ \ \
S_y=\pmatrix{0 & 0 & i \cr 0 & 0 & 0\cr -i & 0 & 0}, \ \ \
S_z=\pmatrix{0 & -i & 0 \cr i & 0 & 0\cr 0 & 0 & 0}.
\label{me6}
\end{equation}
Note that the usual spin commutation relations,
\begin{equation}
\left[S_x,S_y\right]=iS_z,\ \ \
\left[S_y,S_z\right]=iS_x,\ \ \
\left[S_z,S_x\right]=iS_y,
\label{me7}
\end{equation}
hold true as well as
\begin{equation}
S_x^2+S_y^2+S_z^2=S(S+1)\ \ \ {\rm where}\ \ \ S=1.
\label{me8}
\end{equation}
Introducing the complex column three vectors \begin{math} F_+ \end{math}
and \begin{math} F_- \end{math} via
\begin{equation}
F_\pm =\pmatrix{E_x \pm iB_x \cr E_y \pm iB_y \cr E_z \pm iB_z}
\label{me9}
\end{equation}
yields Maxwell's equations in the form
\begin{equation}
i\frac{\partial F_\pm }{\partial t}=\mp ic({\bf S\cdot grad })F_\pm
\label{me10}
\end{equation}
where Eqs.(\ref{me3}), (\ref{me4}) and (\ref{me5}) have been invoked.
Eq.(\ref{me10}) describes (in both righthanded \begin{math} F_+ \end{math}
and lefthanded \begin{math} F_- \end{math} representations) the vacuum field
equations for spin \begin{math} S=1 \end{math} massless photons.

To see more clearly what is involved, let us consider the case of massless
spin one-half particles, i.e. the Weyl massless neutrinos. Defining the momentum
operator as
\begin{equation}
{\bf p}=-i\hbar \ {\bf grad }
\label{me11}
\end{equation}
and with
\begin{math} {\bf \sigma }=(\sigma_x,\sigma_y,\sigma_z) \end{math}
denoting the Pauli matrices,
\begin{equation}
\sigma_x = \pmatrix{0 & 1\cr 1 & 0}, \ \ \
\sigma_y = \pmatrix{0 & -i\cr i & 0}, \ \ \
\sigma_z = \pmatrix{1 & 0\cr 0 & -1},
\label{me12}
\end{equation}
the massless ``Weyl neutrino'' equations read
\begin{eqnarray}
i\hbar \frac{\partial \psi_+}{\partial t} &=&
c({\bf \sigma \cdot p})\psi_+\ \ \ \ \ {\rm righthanded\ Weyl},
\nonumber \\
i\hbar \frac{\partial \psi_-}{\partial t} &=&
-c({\bf \sigma \cdot p})\psi_-\ \ \ \ \ {\rm lefthanded\ Weyl}.
\label{me13}
\end{eqnarray}
The photon analogues to Eqs.(\ref{me13}) are
\begin{eqnarray}
i\hbar \frac{\partial F_+}{\partial t} &=&
c({\bf S \cdot p})F_+\ \ \ \ \ {\rm righthanded\ Maxwell},
\nonumber \\
i\hbar \frac{\partial F_-}{\partial t} &=&
-c({\bf S \cdot p})F_-\ \ \ \ \ {\rm lefthanded\ Maxwell}.
\label{me14}
\end{eqnarray}

There is a difference between the spin one half Eqs.(\ref{me13})
and the spin one Eqs.(\ref{me14}). A lefthanded and a righthanded
neutrino are different particles. Only lefthanded neutrinos
have so far appeared in the laboratory which is a basis for parity
violation. In spite of the wide spread theoretical righthanded cross
product convention for writing down Maxwell's equations,
electromagnetic theory is inherently parity symmetry invariant.
Nevertheless, it is useful to construct a six component field column
vector\cite{Chew:2001} out of the three component righthanded and lefthanded
column vectors defined in Eq.(\ref{me9}); i.e.
\begin{equation}
F=\pmatrix{F_+\cr F_-}.
\label{me15}
\end{equation}
The operators for the photon are now in part described by
\begin{math} 6\times 6 \end{math} matrices naturally constructed
by partitioned \begin{math} 3\times 3 \end{math} sub-matrices.
For example, employing the \begin{math} 3\times 3 \end{math}
spin matrices in Eq.(\ref{me6}) we may write for the
\begin{math} 6\times 6 \end{math} spin one matrices
\begin{equation}
{\bf \Sigma}=\pmatrix{{\bf S} & 0\cr 0 & {\bf S}}.
\label{me16}
\end{equation}
The photon chirality matrix may be defined as
\begin{equation}
\Gamma_5=\pmatrix{1 & 0\cr 0 & -1}.
\label{me17}
\end{equation}
The chiral \begin{math} \Gamma_5 \end{math} spin
\begin{math} {\bf \Sigma } \end{math} matrix product
\begin{equation}
{\bf \alpha }=\Gamma_5{\bf \Sigma}
=\pmatrix{{\bf S} & 0\cr 0 & -{\bf S}},
\label{me18}
\end{equation}
is useful for describing the massless photon Hamiltonian
\begin{equation}
{\cal H}=c({\bf \alpha \cdot p}).
\label{me19}
\end{equation}
Eqs.(\ref{me14}), (\ref{me15}), (\ref{me18}) and and (\ref{me19}) imply the
photon (Maxwell) wave equation in the Schr\"odinger-Dirac form
\begin{equation}
i\hbar \frac{\partial F}{\partial t}={\cal H}F.
\label{me20}
\end{equation}

To obtain the energy eigenvalues for the photon, one needs to solve
\begin{equation}
{\cal H}F_{\bf p}({\bf r})=\tilde{\epsilon }({\bf p}) F_{\bf p}({\bf r}).
\label{me21}
\end{equation}
Choosing
\begin{math} 
F_{\bf p}({\bf r})=F_{\bf p}(0)e^{i{\bf p\cdot r}/\hbar } 
\end{math}
yields the energy eigenvalues
\begin{eqnarray}
\tilde{\epsilon }_+({\bf p}) &=& c|{\bf p}|\ \ \ \ \ \ \ {\rm (photon)},
\nonumber \\
\tilde{\epsilon }_-({\bf p}) &=& -c|{\bf p}|\ \ \ \ \ {\rm (anti-photon)},
\nonumber \\
\tilde{\epsilon }_0({\bf p}) &=& 0 \ \ \ \ \ \ \ \ \ \ \ {\rm (forbidden)},
\label{me22}
\end{eqnarray}
where the anti-photon (negative energy state) is merely the photon moving
backward in time. For \begin{math} {\bf p}\ne 0 \end{math}, the zero 
eigenvalue of energy is {\em forbidden} in virtue of the
vacuum Gauss' law \begin{math} div{\bf E}=0 \end{math} and
\begin{math} div{\bf B}=0 \end{math}.
The photon velocity operator follows from the commutator
\begin{equation}
{\bf v}=\frac{i}{\hbar }\left[{\cal H},{\bf r}\right]
=\frac{\partial {\cal H}}{\partial {\bf p}}
=c{\bf \alpha }=c(\Gamma_5 {\bf \Sigma }).
\label{me23}
\end{equation}
The photon helicity operator \begin{math} \Lambda  \end{math} is
conventionally defined as
\begin{equation}
\Lambda =\left(\frac{\bf p\cdot \Sigma}{|{\bf p}|}\right),
\label{me24}
\end{equation}
The allowed eigenvalues of helicity are
\begin{math} \Lambda=\pm 1 \end{math}
while the eigenvalue \begin{math} \Lambda=0 \end{math}
is forbidden by the Gauss law constraints
\begin{math} div{\bf F}_\pm =0 \end{math} of Eq.(\ref{me3}).
The photon Hamiltonian is given by
\begin{equation}
{\cal H}=({\bf v\cdot p})=
c({\bf \alpha \cdot p})=c\Gamma_5 ({\bf \Sigma \cdot p}),
\label{me25}
\end{equation}
which can then be expressed directly in terms of chirality 
\begin{math} \Gamma_5 \end{math}
and helicity \begin{math} \Lambda \end{math} as
\begin{equation}
{\cal H}=c|{\bf p}|\Gamma_5\Lambda .
\label{me26}
\end{equation}
The eigenvalue spectrum of \begin{math} {\cal H} \end{math} in
Eq.(\ref{me22}) follows directly from the representation in Eq.(\ref{me26}).

\section{Photon Velocity \label{pv}}

Let us now contemplate how fast a photon is moving. For a photon of momentum
\begin{math} {\bf p} \end{math} and energy
\begin{math} \varepsilon ({\bf p})=c|{\bf p}| \end{math}
the mean photon (three vector) velocity is the expected light velocity
value
\begin{equation}
\bar{\bf v}=\overline{\frac{\partial {\cal H}}{\partial {\bf p}}}
=\frac{\partial \varepsilon }{\partial {\bf p}}
=c\left(\frac{\bf p}{|{\bf p}|}\right).
\label{pv1}
\end{equation}
On the other hand, from Eq.(\ref{me23}) we find that photon velocity operator
obeys
\begin{equation}
{\bf v\cdot v}=c^2(\Gamma_5 {\bf \Sigma})\cdot (\Gamma_5 {\bf \Sigma})
=c^2 (\Sigma_x^2+\Sigma_y^2+\Sigma_z^2)=c^2S(S+1)
\label{pv2}
\end{equation}
where \begin{math} \left[\Gamma_5,{\bf \Sigma }\right]=0 \end{math} and
\begin{math} (\Gamma_5)^2=1 \end{math} has been invoked. Since the photon
has spin \begin{math} S=1 \end{math}, it follows that
\begin{equation}
{\bf v\cdot v}=2c^2.
\label{pv3}
\end{equation}
Thus, the root mean square velocity of a photon is the superluminal value
\begin{math} c\sqrt{2} \end{math}. What is the nature of the velocity
fluctuations? The answer resides in the operator for the photon
acceleration \begin{math} {\bf a}=\dot{\bf v}  \end{math}; i.e.
\begin{equation}
\dot{\bf v}=\frac{i}{\hbar }[{\cal H},{\bf v}]=
\frac{i}{\hbar }[({\bf v\cdot p}),{\bf v}].
\label{pv4}
\end{equation}
Employing
\begin{equation}
[v_i,v_j]=c^2[\Gamma_5 \Sigma_i,\Gamma_5 \Sigma_j]=
c^2[\Sigma_i,\Sigma_j]=ic^2\epsilon_{ijk}\Sigma_k
=ic\Gamma_5\epsilon_{ijk}v_k
\label{pv5}
\end{equation}
together with Eq.(\ref{pv4}) yields the equation of motion for 
the photon velocity
\begin{equation}
\dot{\bf v}={\bf \Omega \times v}\ \ \ \ {\rm wherein}
\ \ \ \ \hbar {\bf \Omega}=c\Gamma_5 {\bf p}.
\label{pv6}
\end{equation}
The physical meaning of Eq.(\ref{pv6}) is unambigous. 
The velocity of the photon
precesses about the momentum 
\begin{math} {\bf p}=\hbar {\bf k} \end{math} direction
at an angular velocity
\begin{equation}
{\bf \Omega}_\pm =\pm \left(\frac{c{\bf p}}{\hbar }\right)
=\pm c{\bf k}
\label{pv7}
\end{equation}
with the sign depending on the chirality 
\begin{math} \Gamma_5 \end{math} of the photon.

\begin{figure}[tp]
\centering
\includegraphics[width=2.5in]{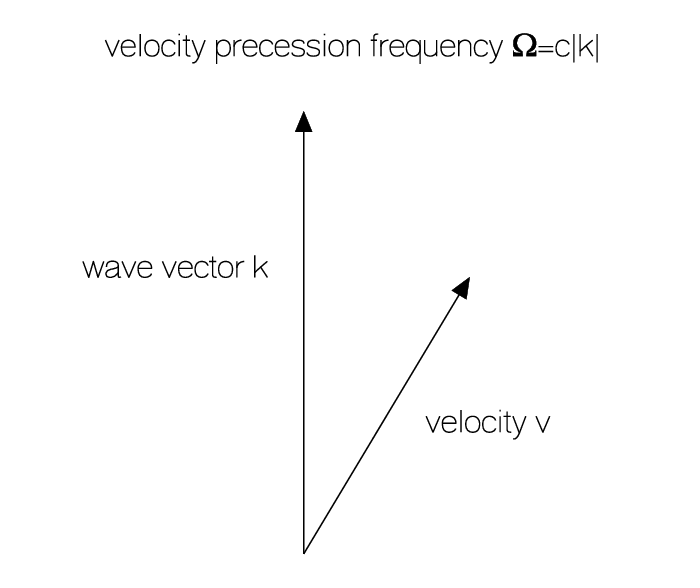}
\caption{The circularly polarized photon velocity ${\bf v}$ precesses
about the wave vector ${\bf k}={\bf p}/\hbar$ axis at an angular velocity
${\bf \Omega }_\pm =\pm c{\bf k}$ with the sign depending on chirality.
The component of the photon velocity parallel to the wave vector
${\bf k}$ is light speed; i.e. $c=({\bf v\cdot k})/|{\bf k}|$.}
\label{fig1}
\end{figure}

\begin{figure}[tp]
\centering
\includegraphics[width=2.5in]{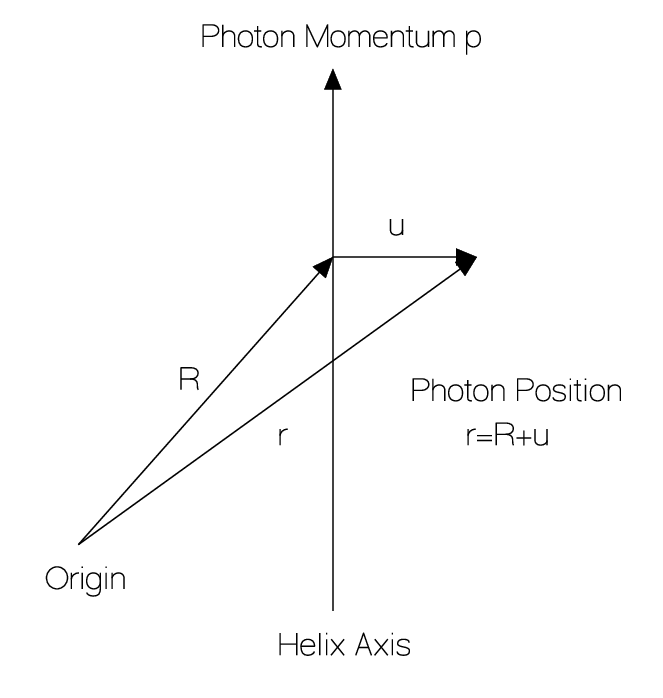}
\caption{For a photon of momentum ${\bf p}=\hbar {\bf k}$, we may
point the momentum along the axis of a helix. The photon position
${\bf r}$ may than be written as a vector ${\bf R}$ pointing to the
helix axis plus a vector ${\bf u}$ normal to the helix axis. The vector
${\bf u}$ moves in a circle at angular velocity ${\bf \Omega}$
which depends upon the chirality via
$\hbar {\bf \Omega}=\Gamma_5c{\bf p}$.}
\label{fig2}
\end{figure}

From the classical electromagnetic theory viewpoint, one envisions either
a right or left circularly polarized electromagnetic wave with frequency
\begin{math} \Omega=c|{\bf k}| \end{math}.
From a photon viewpoint, with \begin{math} {\bf p}=\hbar {\bf k} \end{math},
the photon moves with velocity \begin{math} c \end{math} in the direction of
\begin{math} {\bf k} \end{math}. The photon velocity component perpendicular
to the \begin{math} {\bf k} \end{math}-direction is rotating (in either a 
right or left handed direction) at an angular velocity
\begin{math}{\bf \Omega}_\pm =\pm c{\bf k}\end{math}.
The physical picture is depicted in Fig.\ref{fig1}.

In position space, the photon moves on a ``helical'' path. The velocity 
component along the axis of the helix is light speed. On the other hand, 
the photon is going around and around on a helix with a total 
superluminal speed of \begin{math} c\sqrt{2} \end{math}. 
Let us now consider in more detail how the helical
path enters into the position representation for the photon.

\section{Photon Position and Helicity\label{po}}

Consider the position of the photon as an operator in momentum space
\begin{equation}
{\bf r}=i\hbar \frac{\partial }{\partial {\bf p}}.
\label{po1}
\end{equation}
The above Eq.(\ref{po1}) implies the usual Heisenberg commutation relations
\begin{equation}
[p_k,r_j]=-i\hbar \delta_{jk}.
\label{po2}
\end{equation}
The position 
\begin{math} {\bf r} \end{math} 
of the photon can be decomposed into
a sum of the position \begin{math} {\bf R} \end{math} of the axis of the
helix and the displacement \begin{math} {\bf u} \end{math} from the helix 
axis to the helix coil. As depicted in Fig.\ref{fig2},
\begin{equation}
{\bf r}={\bf R}+{\bf u},
\label{po3}
\end{equation}
where
\begin{equation}
{\bf u}=\frac{\hbar ({\bf \Sigma \times p })}{|{\bf p}|^2}
=\left(\frac{\hbar \Gamma_5}{c} \right)
\frac{\bf v\times p}{|{\bf p}|^2}\ .
\label{po4}
\end{equation}
Radius vector \begin{math} {\bf u} \end{math} rotates in a circle remaining
normal to the photon momentum \begin{math} {\bf p}=\hbar {\bf k} \end{math}
according to the precession equation of motion
\begin{equation}
\dot{\bf u}=\frac{i}{\hbar }[{\cal H},{\bf u}]={\bf \Omega }\times {\bf u}
\ \ \ {\rm wherein}\ \ \ \hbar {\bf \Omega}=c\Gamma_5 {\bf p}.
\label{po4a}
\end{equation}
The radius squared of the helix is then given by
\begin{equation}
{\bf u}\cdot {\bf u}=\left(\frac{\hbar }{|{\bf p}|}\right)^2
\left\{{\bf \Sigma \cdot \Sigma }
-\left(\frac{\bf p\cdot \Sigma}{|{\bf p}|}\right)^2\right\}
=\left(\frac{\hbar }{|{\bf p}|}\right)^2\{S(S+1)-\Lambda^2 \}
\label{po5}
\end{equation}
For a spin \begin{math} S=1 \end{math} photon with helicity
\begin{math} \Lambda =\pm 1  \end{math}, Eq.(\ref{po5}) implies the
square radius of the helix
\begin{equation}
{\bf u}\cdot {\bf u}=\left(\frac{\hbar }{p}\right)^2= \lambdabar^2
\ \ \ ({\rm or\ equivalently}) \ \ \ \sqrt{\bf u\cdot u}=|{\bf
u}|=\lambdabar . \label{po6}
\end{equation}
The radius of the helix orbit is completely determined by the wave length
\begin{math}\lambda =(2\pi \lambdabar)\end{math}.
However, the components of the radius vector from the helix axis
to the helix path do not commute among themselves. In detail,
Eq.(\ref{po4}) implies
\begin{equation}
[u_j,u_k]=
i\left(\frac{\hbar}{|{\bf p}|}\right)^2\Lambda \epsilon_{jkl}
\left(\frac{p_l}{|{\bf p}|}\right) .
\label{po7}
\end{equation}
For example, if a photon moves with a fixed momentum
\begin{math} {\bf p}=(0,0,\hbar /\lambdabar) \end{math} along the
\begin{math}z\end{math}-axis with
helicity \begin{math} \Lambda =1 \end{math}, then the components of
\begin{math} {\bf u} \end{math} do not commute
\begin{math} [u_x,u_y]=i\lambdabar ^2 \end{math}.
Quantum mechanically, the radius of the helix is fixed at
\begin{math} (u_x^2+u_y^2)=\lambdabar ^2 \end{math}
but one cannot tell the values separately of the non-commuting components
\begin{math} {\bf u}=(u_x,u_y,0) \end{math}.

From Eqs.(\ref{pv5}), (\ref{po2}), (\ref{po4}) and (\ref{po7}) 
it follows that
\begin{math}
\left[R_i,u_j\right]=\left[r_i-u_i,u_j\right]=0,
\end{math}
which together with \begin{math} \left[r_i,r_j\right]=0 \end{math} implies
the non-commutative geometry of the coordinates
\begin{math} {\bf R} \end{math}; i.e.
\begin{equation}
i[R_j,R_k]=
\left(\frac{\hbar}{|{\bf p}|}\right)^2\Lambda \epsilon_{jkl}
\left(\frac{p_l}{|{\bf p}|}\right) .
\label{po8}
\end{equation}
If a photon moves with a fixed momentum
\begin{math} {\bf p}=(0,0,\hbar /\lambdabar) \end{math} along the
\begin{math}z\end{math}-axis with
helicity \begin{math} \Lambda =1 \end{math}, then the components of
\begin{math} {\bf R}=(X,Y,Z)  \end{math} do not commute
\begin{math} [X,Y]=-i\lambdabar ^2 \end{math} and the quantum
Pythagorean theorem of Eq.(\ref{intro2}) reads
\begin{math}
\sqrt{X^2+Y^2}\left|n\right>=\lambdabar\sqrt{2n+1}\left|n\right>
\end{math} where \begin{math}n=0,1,2,3,\ldots \ \end{math}.

\section{Photon Angular Momentum\label{am}}
The non-commutative geometry of the Wigner photon coordinate
\begin{math} {\bf R} \end{math} 
can be motivated purely by considerations
relating to the photon angular momentum 
\begin{math} {\bf J} \end{math}.
As is usual one writes the total angular momentum as a sum of the orbital
angular momentum and the spin angular momentum
\begin{equation}
{\bf J}={\bf r\times p}+\hbar {\bf \Sigma}.
\label{am1}
\end{equation}
For a massless particle such as the photon, one seeks to express express
the total angular momentum in terms of helicity. One employs the identity
\begin{equation}
{\bf p\times}({\bf \Sigma \times p})=|{\bf p}|^2
\left\{{\bf \Sigma}-\left(\frac{\bf p}{|{\bf p}|}\right)\Lambda \right\}
\label{am2}
\end{equation}
where the helicity \begin{math} \Lambda \end{math} is defined in
Eq.(\ref{me24}). Eqs.(\ref{am1}) and (\ref{am2}) imply
\begin{equation}
{\bf J}={\bf r\times p}-{\bf u \times p}+
\hbar \left(\frac{\bf p}{|{\bf p}|}\right)\Lambda .
\label{am3}
\end{equation}
where \begin{math} {\bf u} \end{math} is defined in Eq.(\ref{po4}).
If we employ the Wigner photon position
\begin{equation}
{\bf R}={\bf r}-{\bf u}={\bf r}-
\left\{\frac{\hbar ({\bf \Sigma \times p })}{|{\bf p}|^2}\right\},
\label{am4}
\end{equation}
then the total angular momentum may be written in terms of the proper
(zero mass particle) orbital angular momentum
\begin{math} {\bf L}={\bf R\times p} \end{math}
and the helicity \begin{math} \Lambda \end{math} according to
\begin{eqnarray}
{\bf J}&=&{\bf L}+\hbar{\bf \Sigma}_\Lambda \equiv
{\bf R\times p}+\hbar \left(\frac{\bf p}{|{\bf p}|}\right)\Lambda,
\nonumber \\
{\Lambda }&=& \pm 1\ \ \ \ \ \ \ \ ({\rm allowed\ helicity}),
\nonumber \\
{\Lambda }&=& 0 \ \ \ \ \ \ \ \ \ \ \ ({\rm forbidden\ helicity}).
\label{am5}
\end{eqnarray}

It is the Wigner photon coordinate which gives rise to a non-commutative
geometry via
\begin{eqnarray}
[R_j,R_k]&=&-i{\cal A}_{jk}
\nonumber \\
{\cal A}_{jk} &=&
\left(\frac{\hbar}{|{\bf p}|}\right)^2\Lambda \epsilon_{jkl}
\left(\frac{p_l}{|{\bf p}|}\right) \equiv \epsilon_{jkl}A_l.
\label{am6}
\end{eqnarray}
The axial vector area,
\begin{equation}
{\bf A}=\lambdabar^2 \left(\frac{{\bf p}\Lambda }{|{\bf p}|}\right)
\ \ \ {\rm wherein}\ \ \ |{\bf p}|=\frac{\hbar }{\lambdabar},
\label{am7}
\end{equation}
plays an important role in the technology of photon detectors
as will be discussed in what follows. The main physical point is
that the position of a photon in the plain normal to the momentum
is only defined within an area
\begin{math} |{\bf A}|=\lambdabar^2 =|{\bf u}|^2 \end{math}.

\begin{figure}[tp]
\centering
\includegraphics[width=2.5in]{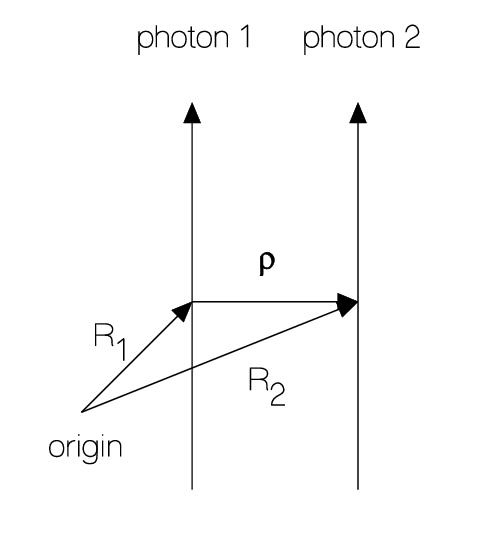}
\caption{Two photons with the same momentum and helicity, move along
different helical paths described by different Wigner coordinate vectors
${\bf R}_1$ and ${\bf R}_2$. The coordinate vector
${\bf \rho}={\bf R}_2-{\bf R}_1$ describes the displacement vector from one
helical axis to the other helical axis.}
\label{fig3}
\end{figure}

\section{Two Photon States\label{pa}}
Consider two photons each with the same momentum 
\begin{math} {\bf p} \end{math} and helicity 
\begin{math} \Lambda \end{math}. Let \begin{math} {\bf R}_1 \end{math}
and \begin{math} {\bf R}_2 \end{math} represent, respectively, 
the Wigner positions of the first and second photon. Finally, let
\begin{equation}
{\bf \rho}={\bf R}_2-{\bf R}_1
\label{pa1}
\end{equation}
represent the vector displacement from the helical axis of the first 
photon to the helical axis of the second photon as shown in 
Fig.\ref{fig3}. From Eqs.(\ref{am6}),
(\ref{am7}) and (\ref{pa1}) it follows that the displacement between 
helical axes obeys the non-commutative geometry rule
\begin{equation}
[\rho_j,\rho_k]=-i\epsilon_{jkl}(A_{12})_l
\ \ \ {\rm with\ area}\ \ \ {\bf A}_{12}=
\left(\frac{\hbar^2{\bf p}_1\Lambda_1}{p_1^3}\right)+
\left(\frac{\hbar^2{\bf p}_2\Lambda_2}{p_2^3}\right).
\label{pa2}
\end{equation}
The distance between the two photon helical axes is then quantized 
in units of \begin{math} \lambdabar \end{math} according to
\begin{eqnarray}
{\bf p}_1={\bf p}_2 &=& (0,0,\hbar /\lambdabar ),
\nonumber \\
(\rho_x^2+\rho_y^2)\left|n\right> &=& 2\lambdabar^2 (2n+1)\left|n\right>,
\nonumber \\
n &=& 0,1,2,3, \ldots \ \ .
\label{pa3}
\end{eqnarray}

Now let us consider a monochromatic beam of photons of a given helicity
incident normal to a plane of photon detectors. Let us also consider
two photon coincidence counts for the photon detectors in the the plane.
According to our central non-commutative geometry Eq.(\ref{pa3}), the Wigner
distance between the two photon coincidence counts should be quantized into
spectrum \begin{math} D_n=\lambdabar \sqrt{2(2n+1)} \end{math} where
\begin{math} n=0,1,2,3,\ldots \ \ \end{math}. Coincidence counting peaks at
detector separations \begin{math} \{D_n\} \end{math} should be observable
in the laboratory if the ``pixel size'' of the photon detectors are small on
the scale of \begin{math} \lambdabar =(\lambda /2\pi) \end{math} where
\begin{math} \lambda \end{math} is the wavelength of the photons.

\section{Conclusion\label{co}}

Rarely in experiments on optics does one try to resolve the photon position
to within a spatial resolution smaller than 
\begin{math} \lambdabar  \end{math}.
For example, in the best commercial digital camera's, the pixel length scale
\begin{math} L > \lambdabar \end{math} since experimentally it appears 
difficult to resolve photon positions on a smaller scale. To the authors' 
knowledge this lower bound on pixel resolution has not been attributed to 
the non-commutative geometry of photon positions but it does appear likely 
that such a geometry sets bounds on how well a photon can be localized.

If the photon detectors are of the electric dipole type
\begin{math} L<< \lambdabar \end{math}, and if two photons in a circularly
polarized beam are simultaneously measured, then the coincidence
count distance \begin{math} D \end{math} can in principle be
measured to a resolution higher than \begin{math} \lambdabar
\end{math}. Observation of peaks spaced at quantized distances
\begin{math} D_n=\lambdabar \sqrt{2(2n+1)} \end{math}
would serve the useful purpose of corroborating the helix picture we have
here theoretically obtained.

The question then is: how easy would it be to obtain experimental
evidence for the predicted effect? Common position-sensitive
devices like charge conductive devices (CCD's) are not 
sensitive to single photons. Devices
suitable for single photon detection based on photocathodes such
as photomultipliers or hybrid photodiodes are unlikely to be
fabricated with pixels small enough to lie inside a wavelength of
light. Nevertheless, there is reason to be optimistic.

Single photon detection  requires a material with a bandgap $E_g$
less than $hc/\lambda$. A useful number to keep in mind is that 1
eV corresponds to a wavelength of 1.2398 microns, which is in the
infrared.

Fabrication requirements would suggest that it makes sense to go
as far into  the infrared as far as possible where there are
essentially two options, both of which are still in rather early
stages of development: 1) superconducting tunnel junctions
(STJ's)\cite{VRP} or 2) pixellated avalanche
photodiodes (APD's)\cite{MRRS}.

Superconducting tunnel junctions have bandgaps of tens of meV,
which correspond to breaking up Cooper pairs. They can, in
principle, be sensitive to  photons of a small fraction of an
electron volt. This would allow pixels of several microns to be
used, but to the best of our knowledge this has not yet been done.

Avalanche photodiodes can be used for single photon counting and
can also be fabricated as finely pixellated devices. Single photon
counting has not yet been demonstrated with position sensitivity,
but should certainly be feasible. Bandgaps are reasonable with
common semiconductors like silicon and germanium having values
less than an electron volt, and thus being sensitive to infrared
of wavelengths longer than a micron. Pixellation at the submicron
level requires masks made using wavelength of light which are much
shorter, but this is in principle possible with ultraviolet light,
or with shorter wavelengths becoming more and more available at
synchrotron light sources.

Quantum efficiencies for both STJ's and APD's can be quite high
(in excess of $60\%$), and noise should not be a problem if one
uses them in a gated mode with a sufficiently intense correlated
photon source. We hope to return to more detailed studies of the
feasibility of an experiment along these lines in the near future.

\end{document}